\documentclass[12pt]{article}

\usepackage{amssymb}
\usepackage{color}
\usepackage{epsfig,amssymb,amsfonts,amsmath,graphicx,cite,xfrac}
\usepackage{authblk}
\usepackage{subcaption}
\definecolor{mygray}{gray}{0.5}

\usepackage{cite}

\newcommand{\be}{\begin{equation}}
\newcommand{\ee}{\end{equation}}
\newcommand{\bea}{\begin{eqnarray}}
\newcommand{\eea}{\end{eqnarray}}

\title{Entanglement of three--qubit random pure states}

\author[${1}$]{Marco Enr\'iquez}
\author[${1}$]{Francisco Delgado}
\author[${2,3}$]{Karol \.Zyczkowski}

\affil[${1}$]{\footnotesize Escuela de Ingenier\'ia y Ciencias, Tecnol\'ogico de Monterrey, Atizap\'an 52926, M\'exico}
\affil[${2}$]{\footnotesize Faculty of Physics, Astronomy and Applied Computer Science, Jagiellonian University, ul. \L ojasiewicza 11, 30-348 Krak\'ow,
Poland}
\affil[${3}$]{\footnotesize Center for Theoretical Physics, Polish Academy of Sciences, Al. Lotnik\'ow 32/46, 02-668 Warsaw, Poland}

\begin{document}

\maketitle
\date{}
\begin{center}
Ver. 4.9
\end{center}
\abstract{We study non-local properties of generic three-qubit pure states. First, we obtain the distributions of both the coefficients and the only phase in the five-term  decomposition of  Ac\'in et al.
for an ensemble of random pure states  generated by the Haar measure on $U(8)$.
Furthermore, we analyze the probability distributions of two sets of polynomial invariants. 
One of these sets allows us to classify three-qubit pure states into four classes. Entanglement in each class is characterized using the minimal R\'enyi-Ingarden-Urbanik entropy. 
Besides, the fidelity of a three-qubit random state with the closest state 
in each entanglement class is investigated. 
We also present a characterization of these classes and the SLOCC classes in terms of the corresponding entanglement polytope.}


\section{Introduction}

Entanglement is possibly the most interesting and complex issue in Quantum Mechanics.
Due to this phenomenon it is not possible to describe properties
of individual subsystems, even though the entire system
is known to be in a concrete pure quantum state.

Quantification of entanglement is still a challenge for any quantum system consisting 
of more than two parts  \cite{WGE16,Ben17}. 
The difficulty of the problem grows fast with 
the growing number of subsystems and becomes intractable 
in the asymptotic limit \cite{gurvits}.
Several measures of quantum entanglement were proposed \cite{HHHH09},
but even in the case of pure states of a multipartie quantum system
it is not possible to identify the single state which can be called
 the most entangled, as the degree of entanglement depends 
 on the measure used  \cite{Enr16}.

On the other hand, 
entanglement in bipartite systems is already well understood. In the case of pure states a key tool in describing non-local properties is the Schmidt decomposition as any entanglement measure is a function of the Schmidt coefficients \cite{Ben17}. Dealing with three-party pure states the problem becomes more intricate as the corresponding state is represented by a tensor rather than a matrix, so one cannot rely 
on the Schmidt decomposition related to the singular value decomposition of a matrix. 
Nevertheless, several decompositions for three-qubit states have been studied in literature 
\cite{Hig00pl, Carteret01,dur}. More recently, a canonical form for symmetric three-qubit states has been proposed showing that in this case the number of entanglement parameters can be reduced from five to three \cite{Mei17}.

Early studies on correlation in composite quantum systems 
revealed that for three or more parties there exist quantum states 
with different forms of entanglement \cite{dur},
as the states from one {\it entanglement class}
cannot be converted by local operations to any states of the other class.
%
As the number of parties increases the number of entanglement classes 
grows fast \cite{versa}.
%
Since local operations cannot generate entanglement, one usually assumes
that a faithful measure of quantum entanglement should be invariant under
local unitary operations and should not grow under arbitrary local operations.
For a given class of operations there exist {\it invariants} 
which are constant along every orbit of equivalent states \cite{albe, grassl}.
A full set of invariants determines a given orbit of locally equivalent states.
However, such sets of invariants 
are established only for systems consisting of few parties of a small dimension
including the simplest multipartite case of three--qubit systems
\cite{sudbery, holweck, Acin01}. 


An interesting question arises, to what extend single-particle properties can provide information about the global entanglement \cite{Saw13}. 
The issue is related to the so-called {\it quantum marginal problem}: 
given a set of reduced density matrices one asks whether they 
might appear as partial trace of a given state of a composed system \cite{Hig03}.
 Necessary conditions for such a 'compatibility problem' were provided in \cite{Bra03}
 for the two-qubit system
 and then developed by Klyachko \cite{Kl04} for the general case. 
 These conditions can be expressed as a set of linear inequalities concerning
 the eigenvalues of the density matrix corresponding to the entire system 
and eigenvalues of the reduced matrices.
 Interestingly, for multipartite systems
 the compatibility problem is related to the entanglement characterization  \cite{Han04}.
 For instance, 
 eigenvalues of three one-qubit reduced matrices of 
 any three-qubit pure state belong to the {\it entanglement polytope} and
 some of its parts correspond to certain classes of quantum entanglement \cite{Wal13}.
 %
%

Not knowing a particular quantum state corresponding to a physical system
it is interesting to ask, what are properties of a typical state.
More formally, one defines an {\sl ensemble} of pure quantum states induced
by the unitary invariant Fubini-Study measure \cite{Ben17}
and computes mean values of various quantities averaging
over the unitary group with respect to the Haar measure.
Such random quantum states are physically interesting
as they arise during time-evolution of quantum systems
corresponding to classically chaotic systems \cite{Kus88,Haa01}
and are relevant for problems of quantum information processing \cite{Zyc05, Gir09}.

Research on non-local properties of generic multipartite  states has been intensive in the last years. This includes entanglement in two qudit systems \cite{Ken02a,Cap06,Kum11}, pairwise entanglement in multi-qubit systems \cite{Viv16,Ken02b,Fac06}, entropic relations and entanglement \cite{Kor14}, correlations and fidelities in qutrits systems \cite{Fan12}, a characterization of entanglement through negativities and tangles in several qubits systems and its relation to the emergence of the bulk geometry \cite{Ran15}.
More recently, genuine entanglement for typical states for a system
composed out of three subsystems with $d$ levels each 
was studied with help of the geometric measure of entanglement \cite{Enr15},
while for generic four-qubit Alsina analyzed the distribution of the hyperdeterminant \cite{Als17}.

The aim of this work is to extend the analysis of
entanglement properties of generic states of three-qubit systems.
We focus our attention on the five-term decomposition of an arbitrary pure state 
\cite{Acin01} as it allows one to construct a set of polynomial invariants 
and to identify the classes of entanglement.
We generated an ensemble of pure quantum states induced
by the Haar measure on the unitary group $U(8)$
corresponding to the system composed of three qubits
and investigated the distribution 
of various entanglement measures and local invariants.


The paper is organized as follows.
In Section 2 we review the five--term decomposition of a three-qubit state
and study statistical properties of the coefficients in such a representation
of a generic state.
In Section 3 we investigate properties the three qubits invariants, $I_k$ and $J_k$ \cite{Acin01}.
We  obtain their probability distributions, either exact or approximate, and
 compare them with accurate numerical approximations. 
Fourth section presents an analysis for the entanglement classes defined in terms of
the latter invariants. 
As a comparative element, we use the R\'enyi and the minimal R\'enyi-Ingarden-Urbanik (RIU) entropies \cite{Enr15} to analyze possible meanings for such classes. Other measure, the maximum overlap with respect to a
selected entanglement class, allows us to identify for an arbitrary three-qubit state the closest state in each class resembling it. 
In Sec. \ref{entpoly} we discuss a characterization of quantum entanglement through the corresponding entanglement polytope and we show how entanglement classes can be distinguished from a geometrical viewpoint.
Last section presents concluding remarks, a list of open questions 
with suggestions concerning the future work.

\section{The canonical five-term decomposition}
A three-qubit state in the Hilbert space ${\cal H}^{\otimes 3}$ involves eight terms, thus, it can be written as
\begin{equation}\label{psi8}
 \vert \psi\rangle = t^{ijk}\vert ijk\rangle, \quad   t^{ijk} {\overline t}_{ijk}=1, \quad t^{ijk}\in \mathbb{C},
\end{equation}
where we have used the repeated indices notation. It  is known \cite{Acin01} that through local unitaries
 the number of terms in $\vert \psi\rangle$ can be reduced from eight to five.
 First, we define the two square matrices $T_0$ and $T_1$ whose entries are given by $(T_i)_{jk}=t^{ijk}$, with $i,j,k=0,1$. A local unitary transformation $U \otimes 1_2 \otimes 1_3$ acting on the first qubit produces
\begin{equation}
 T_0'=u_{00} T_0+u_{01}T_1, \quad T_1'=-{\overline u}_{01} T_0+{\overline u}_{00} T_1.
\end{equation}
The matrix $U$ is taken such that $\det( T_0')=0$. On the other hand, the 
transformation $1_2\otimes V \otimes W$ changes the matrices $T_i$ according to $V T_i W$.
 We choose $V$ and $W$ so that $T_0'$ can be diagonalized via the singular value decomposition (SVD).
 Explicitly, at the end of this procedure we arrive at 

\begin{equation}
 T_0''=\left(\begin{array}{cc} \lambda_0 & 0\\0 & 0\end{array}\right), \quad\quad T_1'' = \left(\begin{array}{cc} {\widetilde \lambda}_1 & {\widetilde \lambda}_2\\{\widetilde \lambda}_3 & {\widetilde \lambda}_4 \end{array}\right).
\end{equation}

In addition, the phase of the coefficients ${\widetilde \lambda}_2, {\widetilde \lambda}_3$ and ${\widetilde \lambda}_4$ can be absorbed into ${\widetilde \lambda}_1$ to yield the decomposition
\begin{equation}\label{acin1}
 \vert \psi \rangle = \lambda_0\vert 000\rangle+\lambda_1e^{i\phi} \vert 100\rangle+\lambda_2\vert 101\rangle+\lambda_3\vert110\rangle+\lambda_4\vert111\rangle,
\end{equation}
where $\lambda_i, \phi \in \mathbb{R}$. Besides $\sum\lambda_i^2=1$. According to \cite{Acin01}, the only phase $\phi$ should be restricted to $0<\phi<\pi$ to assure the uniqueness of the decomposition.

\subsection{Distribution of the coefficients}

\begin{figure}[t] 
\centering
\begin{tabular}{ccc}
\includegraphics[scale=0.57]{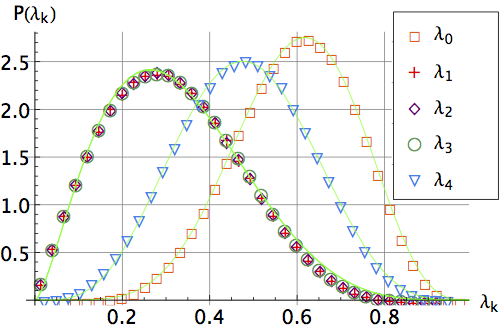} & &
\includegraphics[scale=0.57]{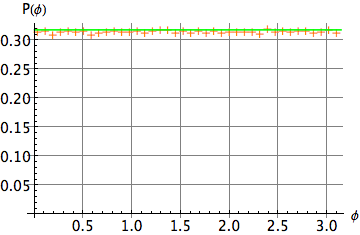}\\
(a) & & (b)

\end{tabular}
\caption{\footnotesize Probability distribution for the Ac\'in coefficients $\lambda_k, k=0,1,...,4$ and the phase $\phi$ in the five-term form (\ref{acin1}) for a set of $10^{6}$ three-qubit random states on ${\mathcal H}_2^{\otimes 3}$. Solid lines represent the best numerical fit in all the cases, the
 parameters of which are listed in Table \ref{bestfitl} }
\label{fig1}
\end{figure}

We take an ensemble of $10^6$ random states in ${\mathcal H}^{\otimes 3}$ distributed according to the unitary invariant measure on the group $U(8)$ and then first reduce them into the five-term representation (\ref{acin1}), then we track each coefficient $\lambda_k$ to compute numerically its probability distributions as well as the distribution of the phase $\phi$. The result is shown in the Figure \ref{fig1} depicting the value of each component $\lambda_k$ versus their relative normalized density 
on ${\mathcal H}^{\otimes 3}$. 
Note that the state (\ref{psi8}) depends on 14 real parameters, say ${\bf p}=(p_0,\ldots,p_{14})$ where we have done the identification $p_\mu = t_{ijk}$, with $\mu=(ijk)_2$. Let us denote by ${\cal P}({\bf t})$ the correspondent joint probability distribution. 
The unitary invariance implies that after the action of the transformation $U\otimes V\otimes W$ on the state $\vert \psi\rangle$ the distribution of the coefficients $\lambda_i$'s and the phase $\phi$ fulfills ${\cal P}({\bf p}) =J\cdot {\cal P}({\bf \lambda})$, where ${\bf \lambda}=(\lambda_0,\lambda_1,\lambda_2,\lambda_3,\lambda_4,\phi)$ and $J$ is the Jacobian of the transformation. The evaluation of this $14\times 14$ determinant becomes cumbersome and 
one has to rely on numerical methods to compute the marginal distributions $P(\lambda_k)$ of the coefficients of the state (\ref{acin1}) as well as the phase $\phi$. The data presented in Fig.~\ref{fig1} (b) suggest that the phase $\phi$ is distributed uniformly on the entire range, $P(\phi)=1/\pi$ for $\phi \in [0,\pi]$.
As the beta distribution has been used to model the behavior of random variables 
limited to finite length intervals in several contexts \cite{Enr15, Gre06, Zyc05}, 
we propose the following distribution $P_i(\lambda_i)=c\, \lambda_i^a (1-\lambda_i)^b$, to fit the distributions of the coefficients $\lambda_j$.
The numerical fits are depicted as solid lines in Fig.~\ref{fig1} (a) and the values of the best fitting parameters are reported in Table~\ref{bestfitl}.
Results presented suggest that the coefficients $\lambda_1$, $\lambda_2$ and $\lambda_3$ are distributed according to the same probability distribution. Hence, we conjecture that out of the six real parameters in
Eq. (\ref{acin1}) only four are required to characterize entanglement in three-qubit random states, say $\{\lambda_0, \lambda_1,\lambda_4, \phi \}$. Interestingly the coefficients $\lambda_0$ and $\lambda_4$ are related with the invariant $J_4$ connected with the three-qubit genuine entanglement 
(for the definition see subsequent section).
As generic three-qubit states are typically strongly entangled \cite{Enr15} 
this analysis illustrates how each coefficient $\lambda_j$ of a given state
is linked with the degree of its entanglement. Note particularly how low values of $\lambda_1, \lambda_2$ and $\lambda_3$ are more representative for entangled states in contrast to $\lambda_4$, the distribution of  which appears to be balanced. Furthermore, the higher values of the coefficient $\lambda_0$
correspond to the states with larger entanglement. This is particularly interesting as in the
decomposition of Carteret {\it et al.} 
this coefficient yields the maximum overlap with the closest separable state \cite{Carteret01}.

\begin{table}[t]
\centering
\begin{tabular}{cccc}
\hline \hline
i & a & b & c\\
\hline
0 & 3.74& 6.05 & 1856.85\\
1 & 67.76 & 4.25 & 1.52\\
2 & 68.40 & 4.27 & 1.53 \\
3 & 66.75 & 4.24 & 1.52 \\
4 & 795.16 & 4.37 & 3.96\\
\hline\hline
\end{tabular}
\caption{\footnotesize Best numerical fit parameters of the distributions $P_i(\lambda_i)=c\, \lambda_i^a (1-\lambda_i)^b$ for $i=0,1,2,3,4$.}
\label{bestfitl}
\end{table}


\section{Three-qubits polynomial invariants}

Local unitary (LU) transformations performed on individual subsystems define 
orbits of locally equivalent multipartite states.
Local invariants can be understood as coordinates in the space of orbits of locally equivalent states.
Any complete set of local invariants allows one to distinguish between 
different orbits of locally equivalent states and thus to describe the degree of 
quantum entanglement \cite{Carteret01}.  For pure states of
a  three--qubit system the space of orbits has six dimensions and it is possible to find 
 six algebraically independent invariants \cite{Carteret02}.

In this section we will analyze the distributions $P(I_k)$ and $P(J_k)$ on ${\mathcal H}^{\otimes 3}$ for the corresponding three-qubit invariants (under local operations) $I_k$ \cite{Sudber01} and $J_k$ \cite{Acin00}, with $k=1,...,5$. These polynomial invariants set representative classes on ${\mathcal H}^{\otimes 3}$ 
and cannot be directly used as the measures of genuine entanglement.

\subsection{Distribution of the invariants}

We first consider the set of five invariants used in \cite{Acin00}:
%
\begin{equation}\label{invac}
\begin{array}{ccc}
I_2={\rm tr}(\rho_A^2), \quad I_3={\rm tr}(\rho_B^2), \quad I_4={\rm tr}(\rho_C^2),\\[1ex]
I_5'''={\rm tr}[(\rho_A \otimes \rho_B)\rho_{AB}], \quad I_6=\vert {\rm Hdet}( T)\vert^2 
\end{array}
\end{equation}
where $\rho_i$ stands for the reduced density matrix of the $i$-th system, 
$\rho_{ij}$ is the reduced density matrix when the partial trace respect the system $k$ is performed 
while $i, j, k$ is a permutation of $A, B, C$. The last invariant is related to the 
hyperdeterminant ${\rm Hdet}$ 
of the tensor coefficients $T=(t^{ijk})$ representing the state (\ref{psi8}).

The invariants are labeled according to the notation used by Sudbery\cite{Sudber01}. 
Note that the squared norm of the state \ref{psi8} is in itself a polynomial invariant 
usually denoted as $I_1$. 
In Figures \ref{invIs} (a-c) we show the probability distribution of the above set of invariants over an ensemble of $10^6$ random states. Moreover, as for $k=2,3,4$ the quantity $I_k$ is related with the linear entropy, $S_k=1-I_k$,  the corresponding distributions show that the entanglement 
of each qubit with the other two is the same no matter which partial trace is performed. 
On the other hand, the invariants $I_k$ in terms of the coefficients $t^{ijk}$ read \cite{Sudber01}:

\begin{figure}[t]
\begin{tabular}{cccc}
\includegraphics[scale=0.6]{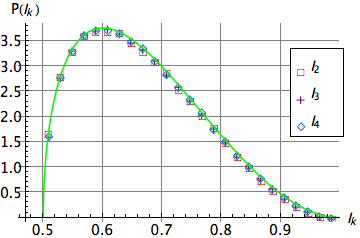} & 
\includegraphics[scale=0.6]{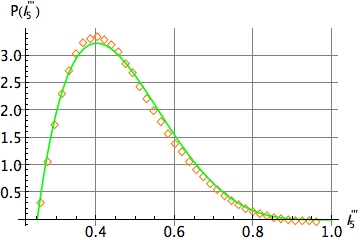} \\
a) & b) \\[1em]
\includegraphics[scale=0.6]{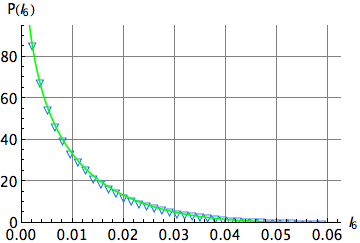} &
\includegraphics[scale=0.33]{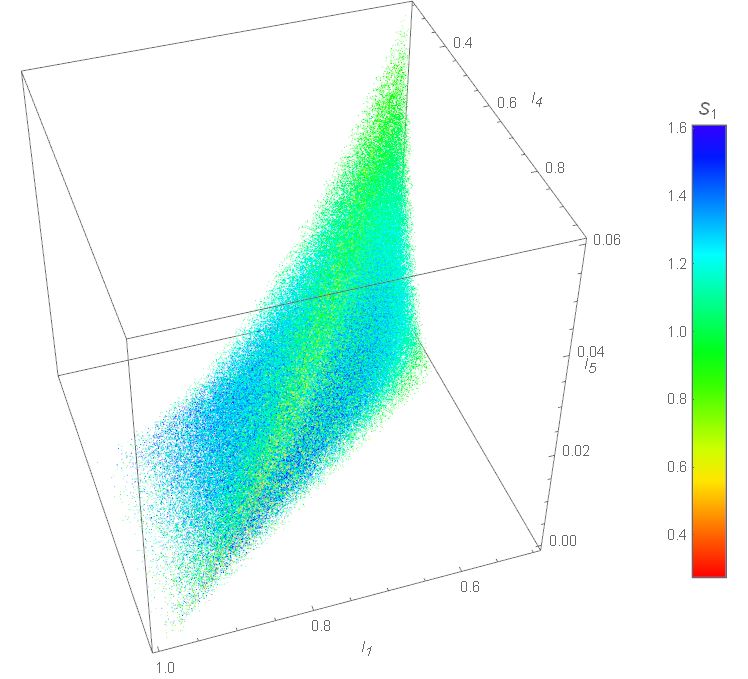}\\
c) & d)
\end{tabular}
\caption{\footnotesize a-c) Probability distribution for the polynomial invariants $I_i, i=1,...,5$ for a set of $10^{6}$ three-qubit random states. Solid line in panel a) stands for the distribution (\ref{PIk}), while  in panels b) and c) the best numerical distributions are depicted by green curves. 
In panel d) a dispersion plot comparing $I_1, I_4$ and $I_5$ is shown.
 In addition, each dot has been colored as function of its $S_1$ R\'enyi entropy \cite{renyi} 
calculated after of the five terms reduction.}
\label{invIs}
\end{figure}
%

\begin{equation*}
\begin{array}{ccc}
\displaystyle I_2= t^{i_1j_1k_1}{\overline t}_{i_2j_1k_1} t^{i_2j_2k_2} {\overline t}_{i_1j_2k_2}, \quad I_3= t^{i_1j_1k_1}{\overline t}_{i_1j_2k_1} t^{i_2j_2k_2} {\overline t}_{i_2j_1k_2}, \quad I_4=\displaystyle t^{i_1j_1k_1}{\overline t}_{i_1j_1k_2} t^{i_2j_2k_2} {\overline t}_{i_2j_2k_1},\\[1em]
I_5'''=t^{i_1j_1k_1}{\overline t}_{i_1j_2k_2} t^{i_2j_2k_2}{\overline t}_{i_2j_3k_1}t^{i_3j_3k_3}{\overline t}_{i_3j_1k_3}\\[1em]
I_6= 4\vert \epsilon_{i_1,j_1} \displaystyle\epsilon_{i_2j_2}\epsilon_{k_1\ell_1}\epsilon_{k_2\ell_2}\epsilon_{i_3k_3}\epsilon_{j_3\ell_3} t^{i_1i_2i_3}t^{j_1j_2j_3}t^{k_1k_2k_3}t^{\ell_1\ell_2\ell_3} \vert^2
\end{array}
\end{equation*}
where the convention of summation over repeated indexes is used 
and $\epsilon_{k,\ell}$ stands for the Levi-Civita tensor of order two. Since the coefficients can be regarded as a column of a random unitary matrix, we can compute the average value of each invariant
 by evaluating integrals of polynomial functions over the unitary group with 
respect to unique normalized Haar measure. 
Using using symbolic integration \cite{Puc17} we obtain $\langle I_k\rangle=2/3$ for $k=2,3,4$. This result is consistent with the mean purity of a single qubit traced out from a $2\times 4$ system reported in \cite{Zyc01}. Moreover, $\langle I_5'''\rangle=7/15$ and  $\langle I_5'''^2\rangle=133/572$. 
In order to compute the mean value of $I_6$ we use the second moment of the three-tangle $\tau$ reported in \cite{Enr15} with the fact $\tau^2 =16 I_6$ to get $\langle I_6\rangle = 1/110$. 
On the other hand, to compute the distributions of the invariants $P(I_k)$ for $k=2,3,4$ we first note that the joint density of eigenvalues $\vartheta_1$ and $\vartheta_2$ of a single qubit traced out of a system of a three-qubit system is given in equation (3.6) of \cite{Zyc01}
with $N=2$ and $K=4$. This reads
\begin{equation}\label{plambdas}
 {\cal P}(\vartheta_1,\vartheta_2)=210\, \delta(1-\vartheta_1-\vartheta_2) (\vartheta_1-\vartheta_2)^2 \vartheta_1^2\vartheta_2^2
\end{equation}
where $\delta$ stands for the Dirac delta. As each $I_k$ is nothing else than the purity of a single qubit reduced density matrix, we can compute the probability distribution by performing the following integral
\begin{equation}
 P(I_k)=210 \int_0^1\int_0^1 d \vartheta_1 d\vartheta_2 {\cal P}(\vartheta_1,\vartheta_2) \delta(I_k-\vartheta_1^2-\vartheta_2^2),
\end{equation}
this yields
%
\begin{equation}\label{PIk}
 P(I_k)=\frac{105}2 (1-I_k)^2 (2 I_k-1)^{1/2}, \quad 1/2\le I_k \le1, \quad k=2,3,4.
\end{equation}
This probability distribution is depicted in Fig.~\ref{invIs}. In addition, we approximate the distribution $P(I_5''')$ by the following beta distribution
\begin{equation}\label{PF4}
 P_{F_5}(I_5''')= \frac{\Gamma(a+b+2)}{3^{a+b+1} 4^{a+1} \Gamma(a+1)\Gamma(b+1)} (1-I_5''')^a (4I_5'''-1)^b, 
\end{equation}
requiring the first two moments of this distribution coincide with the exact two moments of $P(I_5''')$ reported above. We found $a=21989/5691$ and $b=5554/5691$. On the other hand, the distribution of the square of the three tangle was approximated in \cite{Enr15} by a Beta distribution. Thus, making a variable change in this result we may approximate $P(I_6)$ by
\begin{equation}\label{pfI6}
 P_{F_6}(I_6) = \frac 2{\sqrt{I_6}} {\rm Beta}(31/17, 62/17, 4 \sqrt{I_6}), \quad 0\le I_6\le 1/16.
\end{equation}
As the distributions of the invariants $I_2, I_3$ and $I_4$ are the same, thus we only need three invariants to characterize the entanglement in the set of three-qubit random states, say ${(I_2,I_5''',I_6})$. In Figure \ref{invIs}d) we show a dispersion plot whose three axes correspond to such invariants and their colors correspond to their $S_1$ R\'enyi entropy calculated after of the five terms reduction \cite{Enr15} (which will be properly presented in the next section) in agreement with the side color scale.

\begin{figure}[t]
\begin{tabular}{cccc}
\includegraphics[scale=0.6]{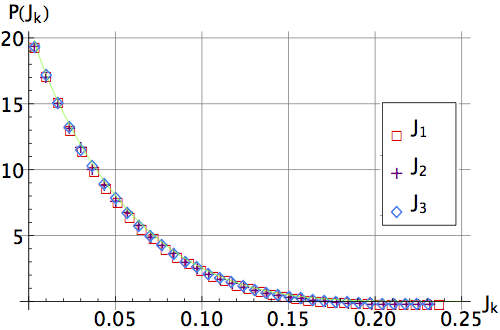} & 
\includegraphics[scale=0.6]{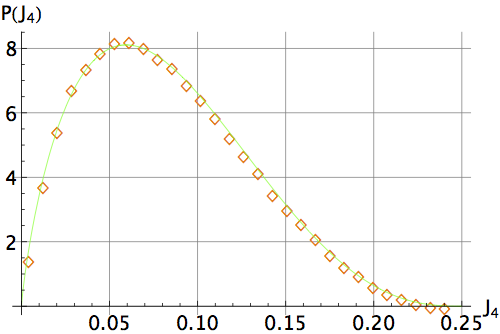} \\
a) & b) \\[1em]
\includegraphics[scale=0.6]{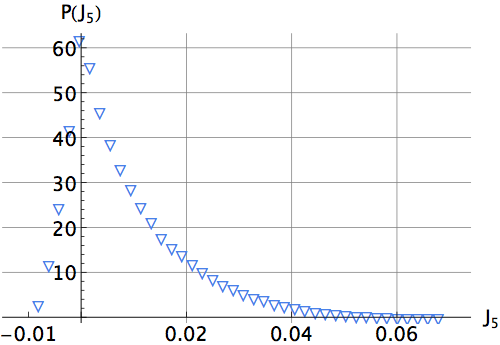} &
\includegraphics[scale=0.33]{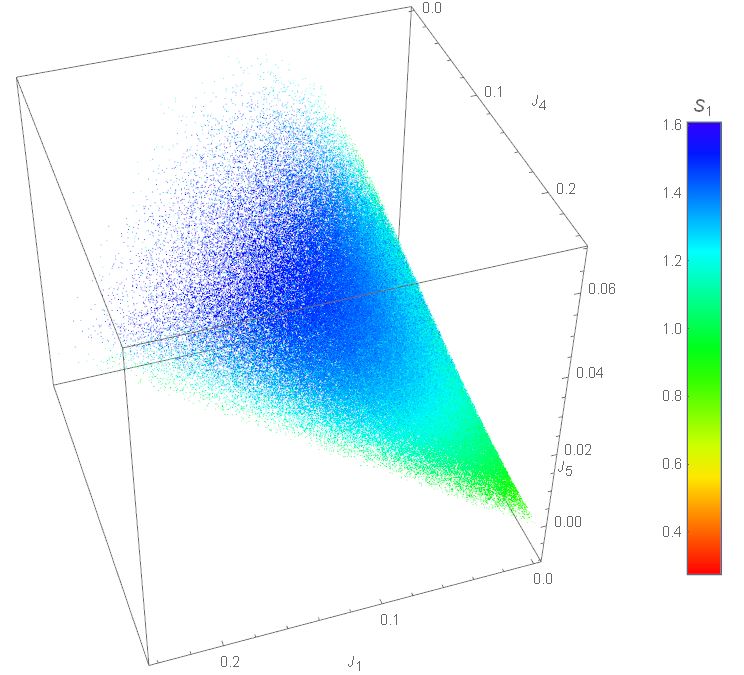}\\
c) & d)
\end{tabular}
\caption{\footnotesize a-c) Probability distribution for the polynomial invariants $J_i, i=1,...,5$ for a set of $10^{6}$ three-qubit random states. In all graphics the numerical best fit distribution is depicted as the green line. In Fig. d) we show a dispersion plot comparing $J_1, J_4$ and $J_5$. In addition, each dot has been colored as function of its $S_1$ R\'enyi entropy \cite{renyi} calculated after of the five terms reduction in agreement with the side color scale.}
\label{invJs}
\end{figure}

We also consider the set of invariants proposed by Ac\'in {\it et al.} \cite{Acin01}. These invariants allow to identify different entanglement classes (which will be discussed in next section) and can be written in terms of the six parameters of the five-term decomposition as
\begin{equation}
\begin{array}{ccc}
J_1=\vert \lambda_1\lambda_4 e^{i\varphi}-\lambda_2\lambda_3\vert^2, \quad J_2=\mu_0\mu_2, \quad J_3=\mu_0\mu_3,\\[1ex]
J_4=\mu_0\mu_4, \quad J_5 = \mu_0(J_1+\mu_2\mu_3-\mu_1\mu_4),
\end{array} 
\end{equation}
where $\mu_i=\lambda_i^2$. For this analysis, the same set of $10^6$ random states was considered but now obtaining the corresponding values of them through their expressions in terms of the five-term coefficients \cite{Acin01}. All these invariants can be calculated departing from the set of $\lambda_i$. The outcomes are shown in the Figures \ref{invIs} and \ref{invJs} in their respective ranges. Note in the Figure \ref{invJs}a-c how for $J_1, J_2$ and $J_3$ the distribution is biased on low values of these invariants, denoting a possible relation with higher entanglement. For $J_4$, properly the Hyperdeterminant, the distribution peaks around of $\frac{1}{16}$, denoting that separability as well as genuine entanglement are absent in the most of states in ${\mathcal H}^{\otimes 3}$. A similar feature is observed for $J_5$ but varying sharply for negative and positive values. On the other hand, the invariants $J_k$'s can be expressed in terms of $I_k$'s \cite{Acin01}:
\begin{equation*}
\begin{array}{cc}
 J_1=\displaystyle\frac14(1+I_2-I_3-I_4-2\sqrt{I_6}), \quad J_2=\displaystyle\frac14(1-I_2+I_3-I_4-2\sqrt{I_6}), \\ 
J_3=\displaystyle\frac14(1-I_2-I_3+I_4-2\sqrt{I_6}), \quad J_4=\sqrt{I_6}, \\
J_5=\displaystyle\frac14(3-3I_2-3I_3-I_4+4I_5-2 \sqrt{I_6}).
\end{array}
\end{equation*}
Such expressions are useful to compute some averages. For instance, as $\langle \sqrt{I_5}\rangle = \langle \tau \rangle/4$ it is immediate to compute $\langle J_4\rangle = 1/12$. From the above definitions we can calculate directly $\langle J_k\rangle = 1/24$, for $k=1,2,3$ and $\langle J_5\rangle = 1/120$. We approximate the probability distributions $P(J_k)$ with $k=1,2,3$ by a distribution $P_{F_k}(J_k)\sim J_k^a (1-4J_k)^b$, where the parameters in this case are determined numerically to yield the best fit.
In addition, making use of the approximation (\ref{pfI6}) for the distribution of the 
invariant $I_6$
one can obtain the following approximation for the distribution of the variable $J_4$
\begin{equation}
 P_{F4}(J_4)=4 {\rm Beta}(31/17,62/17; 4 J_4), \quad 0\le J_4\le 1/4
\end{equation}
%
%
%
%
 On the other hand, as the distributions for $J_1,$ $J_2$ and $J_4$ are uniform among them, we may characterize the entanglement using only the invariants $J_1, J_4 $ and $J_5$. In Figure \ref{invJs}d we depict a scatter plot using these invariants as coordinates similarly as in the Figure \ref{invIs}d for $I_k$. 

\begin{figure}[t]
\centering
\includegraphics[scale=0.6]{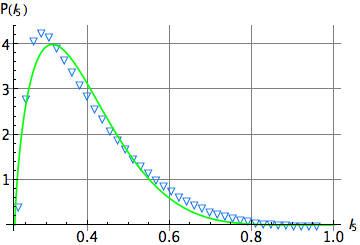}\hspace{1pc} 
\caption{\footnotesize Probability distribution of Kempe invariant $I_5$ obtained using $10^6$ random states. The green line stands for the probability distribution (\ref{fitk}).}
\label{distk}
\end{figure}

Another interesting invariant is the one obtained by Kempe \cite{Kempe01}:
\begin{equation}\label{invk}
 I_5=3 {\rm tr} (\rho_A\otimes \rho_B)\rho_{AB}-{\rm tr} \rho_A^3-{\rm tr}\rho_B^3=t^{i_1j_1k_1}t^{i_2j_2k_2}t^{i_3j_3k_3}{\overline t}_{i_1j_2k_3}{\overline t}_{i_2j_3k_1} {\overline t}_{i_3j_1k_2},
\end{equation}
which distinguishes locally indistinguishable states. In terms of the Ac\'in parameters it 
reads
\begin{eqnarray}
I_5 &=& 1 -3 \lambda_4^2 -3 \lambda_3^2 +3 \lambda_3^4+3 \lambda_4^4+3 \lambda_1^2 \lambda_3^2+6 \lambda_3^2 \lambda_4^2 \\
&& \quad +\left(\lambda_1^2 \left(3-6 \lambda_3^2\right)-3 \left(\lambda_3^2-1\right) \left(2 \lambda_3^2+2 \lambda_4^2-1\right)\right) \lambda_2^2 \nonumber \\ 
&& \quad + 6 \lambda_1 \lambda_3 \lambda_4 \left(\lambda_1^2+\lambda_2^2+\lambda_3^2+\lambda_4^2\right) \lambda_2 \cos \phi +\left(3-6    \lambda_3^2\right) \lambda_2^4. \nonumber 
\end{eqnarray}
%
%
Note that the form  (\ref{invk}) of the Kempe Invariant $I_5$ is manifestly  permutation symmetric.
Although this quantity cannot be considered as a legitimate measure of entanglement,
Osterloh has pointed out  \cite{Ost10} that different values of $I_5$ allow to distinguish between
different local orbits of three qubit pure states.
Integrating equation (\ref{invk}) using symbolic integration on the Haar measure, we found that $\langle I_5\rangle = 2/5$ and $\langle I_5^2\rangle=499/2860$. In Figure \ref{distk} we show the probability distribution of the invariant $I_5$, which can be approximated by the distribution
\begin{equation}\label{fitk}
 P_{F_{I_5}}(\kappa)=\frac{9^{a+1}\Gamma(a+b+2)}{7^{a+b+1} \Gamma(a+1)\Gamma(b+1)}(1-\kappa)^a(9\kappa-2)^b, \quad 2/9\le\kappa\le1
\end{equation}
where $a=90/23$ and $b=283/621$ are set by the condition that the first two moments of $P_{F_{I_5}}(I_5)$ correspond with the first two moments of $P(I_5)$ provided above. Remark that  sextic invariant $I_5'''$ can be written in terms of the Kempe invariant and the quadratic and quartic invariants \cite{Sudber01}

\section{Three-qubits entanglement classes}

A state classification has been presented in \cite{Acin01} based on the minimal number of product states in (\ref{acin1}). Ac\'in et al. reported some entanglement classes wich are presented in Table ~\ref{tab1}.
The conditions for such classes states are expressed in terms of the invariants $J_k$. 
Thus, in this section we consider the invariant classes introduced there, departing from the coefficients of the five-term representation in ${\mathcal H}^{\otimes 3}$. These classes barely describe some families around some characteristic states in this space. The first aim is to analyze how those classes represent the entanglement of each state included there, mainly based on the entanglement distribution knowledge on ${\mathcal H}^{\otimes 3}$ \cite{Enr15}. 
%
%
%
%
%
%
Note that in some classes the direct imposition of the conditions on the invariants leaves some product states that differ from those reported by Ac\'in, that is to say, to obtain such product states an additional LU transformation is required. Such cases are remarked with $\star$ in Table ~\ref{tab1}.
\begin{table} 
\fontsize{10pt}{4}\selectfont
\centering
\begin{tabular}{l p{4cm} lll}
\hline\hline
Class & Conditions & States & Entanglement polytope\\
\hline\\[1ex]

1 & $J_i=0$ & $\vert 000\rangle$ & point ${\mathcal O}=(0,0,0)$\\[1em]
2a & All $J_i=0$ apart from $J_1$ & $\vert 000\rangle, \vert 011\rangle^\star$ &  lines $\overline{{\mathcal O}A}$, $\overline{{\mathcal O}B}$ and $\overline{{\mathcal O}C}$\\[1em]
2b & All $J_i=0$ apart from $J_4$ & $\vert 000\rangle, \vert 111\rangle$ & line $\overline{{\mathcal O}G}$ \\[1em]
3a & $J_1J_2+J_1J_3+J_2J_3=\sqrt{J_1J_2J_3}=J_5/2$, $J_4=0$ & $\vert 000\rangle, \vert 101\rangle, \vert110\rangle$ & $\triangle_2 {\mathcal O} AB$, $\triangle_2 {\mathcal O} AC$, $\triangle_2 {\mathcal O} BC$, $\triangle_2 ABC$\\[1em]
3b & $J_1=J_2=J_5=0$ & $\vert 000\rangle, \vert 110\rangle, \vert111\rangle$ & $\triangle_2 ABG$, $\triangle_2 ACG$, $\triangle_2 BCG$\\[1em]
4a & $J_4=0, \sqrt{J_1J_2J_3}=J_5/2$ & $\vert000\rangle, \vert100\rangle, \vert101\rangle,\vert110\rangle$ & $\triangle_3 {\mathcal O}ABC$\\[1em]
4b & $J_2=J_5=0$ & $\vert000\rangle, \vert100\rangle, \vert110\rangle,\vert111\rangle$\\[1em]
4c & $J_1J_4+J_1J_2+J_1J_3+J_2J_3=\sqrt{J_1J_2J_3}=J_5/2$ & $\vert000\rangle, \vert101\rangle, \vert110\rangle,\vert111\rangle$\\[1em]
4d & $\Delta_J=0, \sqrt{J_1J_2J_3}=\vert J_5\vert/2$ & $\vert000\rangle, \vert010\rangle, \vert100\rangle,\vert111\rangle^\star$\\[1em]
\hline\hline
\end{tabular}
\caption{\footnotesize Ac\'in entanglement classes introduced in \cite{Acin01}.
Besides $ \mbox{\scriptsize $\Delta_J\equiv (J_4+J_5)^2-4(J_1+J_4)(J_2+J_4)(J_3+J_4)$} $. 
Basis elements marked with $\star$ are not directly obtained instead they have additional relabellings. Besides, fourth column shows the identification of each class with subsets of the entanglement polytope. The point $G$ stands for $(1/2,1/2,1/2)$. Details are presented in Section \ref{entpoly}. }
\label{tab1}
\end{table}

\subsection{The minimal decomposition entropy}
We characterize the entanglement degree of the classes in Table \ref{tab1} using the minimal R\'enyi-Ingarden-Urbanik (RIU) entropy also known as minimal decomposition entropy \cite{Enr15}. For the state (\ref{psi8}) this is defined as
\begin{equation} 
S^{\rm RIU}_q \left(\psi \right) \ := \
 \min_{U_{\rm loc}}   S_q \left[ p( U_{\rm loc} \vert \psi\rangle )\right],
\label{RIU}
\end{equation}
where $p(\cdot)$ stands for the probability vector related to the state (\ref{psi8}) and the minimum is taken on all local transformations $U_{\rm loc}=U_1\otimes U_2\otimes U_3$. Remark that $S_q$ is the $q-$order R\'enyi entropy \cite{renyi}. Depending on the parameter $q$ the quantity (\ref{RIU}) provides information about the state \cite{Enr15}. Thus, for
\begin{itemize}
\item $q=0$: the decomposition entropy is related to the tensor rank of the
 state $\vert \psi\rangle$. As a direct consequence of the decomposition  (\ref{acin1})
we have $S_0^{\rm RIU}(\psi)\le 5$.
\item $q=1$: the minimal decomposition entropy $S^{\rm RIU}_1 (\vert \psi\rangle)$ determines the minimal
 information gained by the environment after performing a projective von--Neumann
 measurement of the pure state $\vert \psi\rangle \langle \psi \vert$ in an 
arbitrary product basis \cite{Maz01}.
\item $q\rightarrow \infty$: in such limiting case the minimal RIU entropy is associated with the maximal  overlap with the closest separable state $\Lambda_{\max}=\max \vert \langle \psi \vert \chi_{\rm sep}\rangle \vert ^2$. Indeed, it can be shown that $S^{\rm RIU}_{\infty} (\vert \psi\rangle)= - \log \lambda_{\rm max}$. See \cite{Enr15} for details.
\end{itemize}
A direct computation shows that for a state in class 1 the minimal RIU entropy vanishes regardless the value of the parameter $q$. 
The corresponding calculation for the other entanglement classes is presented below.
\subsubsection{Classes 2}
A direct calculation shows that the decomposition of states in class 2b is optimal.
That is to say, if the state is given by
\begin{equation}
 \vert \varphi_{\rm 2b} \rangle = \cos \alpha \vert000\rangle+\sin\alpha \vert 111\rangle, \quad 0< \alpha < \pi/2,
\end{equation}
the minimal decomposition entropy reads
\begin{equation}
 S_1^{\rm RIU} (\varphi_{2b})=-\cos^2\alpha \ln (\cos^2\alpha)-\sin^2\alpha \ln (\sin^2\alpha).
\end{equation}
Our numeric calculations indicate that for the class 2a the Ac\'in decomposition is optimal as well. 
The states with the largest minimal decomposition entropy in each class are
\begin{equation}
 \vert \varphi_{\rm 2a}^{\rm max}\rangle=\frac1{\sqrt 2} \vert000\rangle+\frac1{\sqrt 2}\vert 111\rangle, \quad \vert \varphi_{\rm 2b}^{\rm max}\rangle=\frac1{\sqrt 2} \vert100\rangle+\frac1{\sqrt 2}\vert 111\rangle,
\end{equation}
note the reported basis for class $2$b in Table \ref{tab1} is different due to additional changes commonly reported in the literature. A simple calculation shows the LU equivalence of the two local basis. Note that the state $\vert \varphi_{\rm 2b}^{\rm max}\rangle$ is bi-separable and it 
attains the same 
minimal decomposition entropy as the GHZ state.

\subsubsection{Classes 3}
%
Any state belonging to class 3a can be parametrized as
\begin{equation*}
 \vert \varphi_{\rm 3a} \rangle = \sin\theta_1\sin \theta_2 \vert000\rangle +\sin\theta_1\cos\theta_2 \vert 101\rangle+\cos\theta_1\vert 110\rangle, \quad 0<\theta_1,\theta_2<\pi/2 
\end{equation*}

\noindent Note such  state is LU-equivalent to the symmetric state
\begin{equation}\label{symst3a}
 \vert {\widetilde \varphi}_{\rm 3a} \rangle = \sin\theta_1\sin \theta_2 \vert100\rangle +\sin\theta_1\cos\theta_2 \vert 001\rangle+\cos\theta_1\vert 010\rangle, 
\end{equation}
hence, the minimal RIU entropy can be computed using the method described in \cite{Enr15} for symmetric states. In particular if $\cos \theta_1=1/\sqrt3$ and $\sin\theta_2=1/\sqrt 2$ we obtain the well-known W-state for which $S_1^{\rm RIU}(W)=\ln 3$, which is the largest value of $S_1^{\rm RIU}$ for this class.

\begin{figure}[h] 
\centering
\begin{tabular}{ccc}
\includegraphics[scale=0.45]{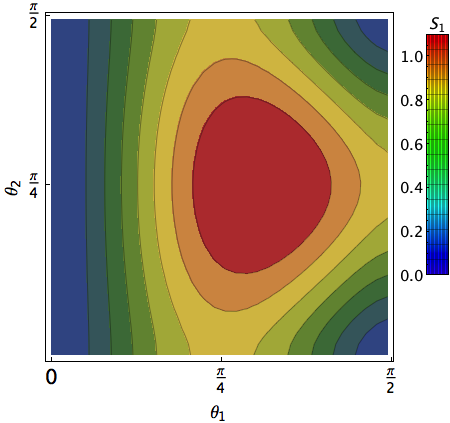} & \quad &\includegraphics[scale=0.45]{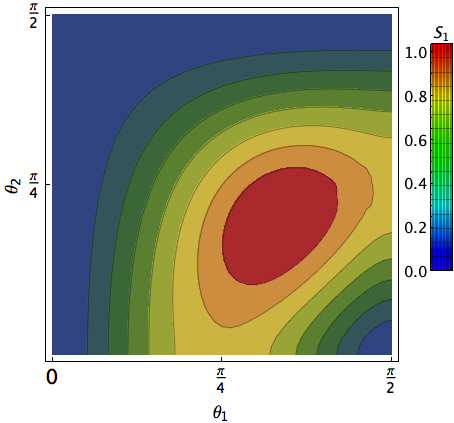} \\
a) & \qquad & b)
\end{tabular}
\caption{\footnotesize The minimal decomposition entropy level curves for classes a) 3a, and b) 3b.}
\label{s1c3}
\end{figure}

On the other hand, a state in class 3b can be written as
\begin{equation*}
 \vert \varphi_{\rm 3b} \rangle = \sin\theta_1\sin \theta_2 \vert000\rangle +\sin\theta_1\cos\theta_2 \vert 110\rangle+\cos\theta_1\vert 111\rangle, \quad 0<\theta_1,\theta_2<\pi/2.
\end{equation*}
No state in class 3b has greater $S_1^{\rm RIU}$ than the $W$-state. For a general state in these classes the minimal decomposition entropy as function of parameters $\theta_1$ and $\theta_2$ is depicted in Fig. \ref{s1c3}. Note that regions of maximal $S_1^{\rm RIU}$ entropy are around the values $\theta_1, \theta_2$ for the maximal entropy for such states. 
\subsubsection{Classes 4}
\begin{figure}[ht] 
\centering
\begin{tabular}{ccc}
\includegraphics[scale=0.41]{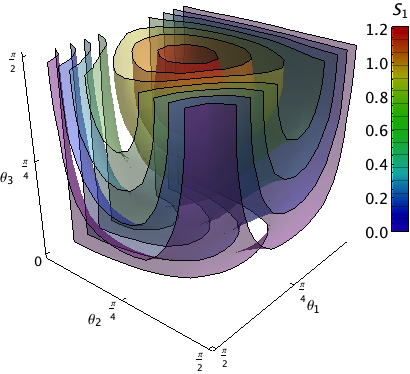} & \quad &\includegraphics[scale=0.36]{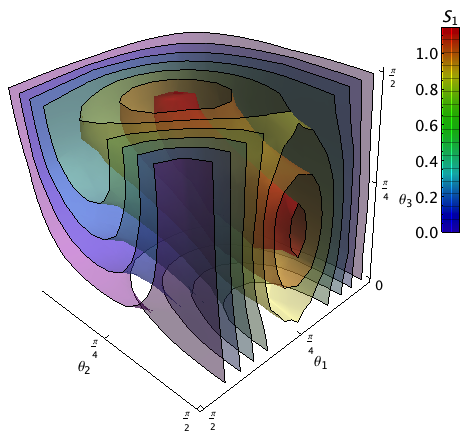}\\
a) & \quad & b)\\[1em]
\includegraphics[scale=0.35]{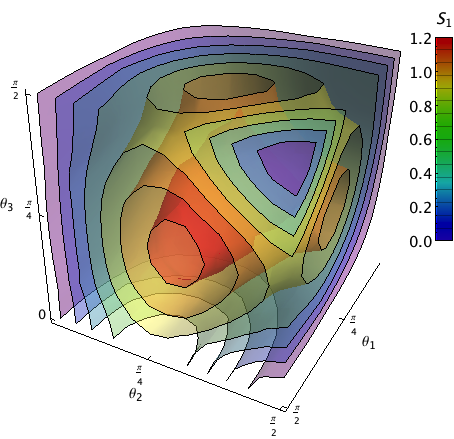} &\quad&\includegraphics[scale=0.36]{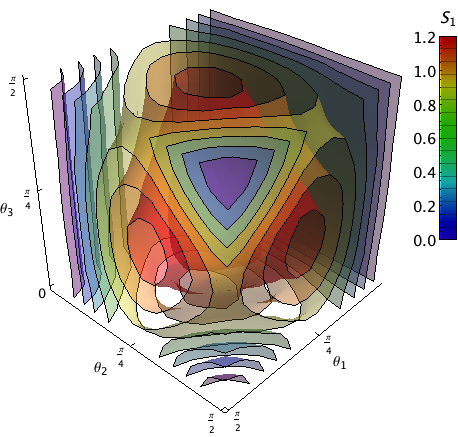} \\
c) & \quad & d)
\end{tabular}
\caption{\footnotesize Surfaces of equal entanglement for classes 4 measured with respect the minimal decomposition entropy.}
\label{s1c4}
\end{figure}

A general state in each one of the classes 4 can be written as
\begin{eqnarray}
\vert \varphi_{\rm 4a}\rangle &=& \beta_1 \vert 000\rangle+ e^{i\varphi} \beta_2 \vert 100\rangle+ \beta_3 \vert101\rangle+ \beta_4 \vert 110\rangle \\
\vert \varphi_{\rm 4b}\rangle &=& \beta_1 \vert 000\rangle+ e^{i\varphi} \beta_2 \vert 100\rangle+ \beta_3 \vert110\rangle+ \beta_4 \vert 111\rangle \\
\vert \varphi_{\rm 4c}\rangle &=& \beta_1 \vert 000\rangle+ \beta_2 \vert 101\rangle+ \beta_3 \vert110\rangle+ \beta_4 \vert 111\rangle \\
\vert \varphi_{\rm 4d}\rangle &=& \beta_1 \vert 000\rangle+ \beta_2 \vert 010\rangle+ \beta_3 \vert100\rangle+ \beta_4 \vert 111\rangle
\end{eqnarray}

\noindent where $\beta_1=\sin \theta_0\sin\theta_1\sin\theta_2, \beta_2=\sin\theta_0\sin\theta_1 \cos\theta_2, \beta_3=\sin\theta_0\cos\theta_1$ and $\beta_4=\cos\theta_0$. As for class $2$b, the basis elements for class $4$d reported in Table \ref{tab1} are not the directly obtained from (\ref{acin1}). Class $4$b correspond to the real class (with all components real, thus $e^{i \varphi}=\pm 1$) which lets to perform an additional reduction to only four terms. As in the previous case, we get the surfaces of minimal decomposition entropy in terms of parameters $\theta_0, \theta_1$ and $\theta_2$ in the Figure \ref{s1c4}. Those figures exhibit for each class the behavior for the  entropy. There, the frontiers of the regions shown $\theta_1, \theta_2, \theta_3 = 0, \pi/2$ correspond to separable states. In addition, our numerical calculations show that the minimal decomposition entropy is independent of the phase $\phi$. We also numerically found that the the largest $S_1^{\rm RIU}(\psi_{\rm 4a}^{\rm max})= 1.213$ is attained for a state in class 4a with $\theta_1=3\pi/10$, $\theta_2=4\pi/15$ and $\theta_3=23\pi/60$. Note
that this value is smaller than the one reported earlier \cite{Enr15}
 as the maximal for a random state with five components.

\subsection{The maximum overlap with an entanglement class}

Given an ensemble of random states, a natural question arises: how many states of such ensemble belong to a particular Ac\'in entanglement class? To tackle  this question observe first that 
numeric calculations imply $\langle S_0^{\rm RIU} (\psi)\rangle=\log 5$. Hence a generic three-qubit state has five non trivial components in the decomposition (\ref{acin1}). 
As each class has at most four components, we rather consider the following quantity
\begin{equation}\label{projection}
 \Lambda_i(\beta) = \max_{\vert\varphi\rangle , U_{\rm local}} \{\vert \langle \varphi \vert U_{\rm local}^\dagger \vert \beta \rangle\vert^2: \vert \varphi \rangle \in \mbox{ Class }i \},
\end{equation}
\noindent where $i =\{ 1, 2a, 2b, 3a, 3b, 4a, 4b, 4c, 4d \}$ and $U_{\rm local} = U_1 \otimes U_2 \otimes U_3$. Such quantity provides an information, how much a given state $\vert\beta\rangle$ on ${\mathcal H}^{\otimes 3}$ differs from the closest state $\vert \varphi \rangle$ in the Ac\'in entanglement class $i$ \cite{Acin01}.
Note that the quantity $\Lambda_i$ can be interpreted as the maximal fidelity 
of a given state $|\beta\rangle$ with respect to the closest state belonging to the class $i$. 
In particular, if $i=1$ the results are consistent with $S_\infty^{RIU}(\beta)$ (see \cite{Enr15}) as this yields the maximum overlap with the closest separable state.

By taking a set of $10^5$ random states in ${\mathcal H}^{\otimes 3}$, we get their projection $\Lambda_i$ on each Ac\'in class, tracking their Hyperdeterminant ${\rm Hdet}(\vert \varphi \rangle)$, which is clearly invariant under local transformations. Then we perform a numerical optimization on the three parameters depicting a local transformation on each qubit (nine as total) together with the necessary coefficients depicting an arbitrary state in each class \cite{Acin01}. Finally, we track also the Hyperdeterminant of such state, ${\rm Hdet}(\vert \beta \rangle)$. With this information we construct the corresponding distribution $\rho(\Lambda_i)$ of each projection $i$ (\ref{projection}).

\begin{figure}[t] 
\centering
\includegraphics[width=32pc]{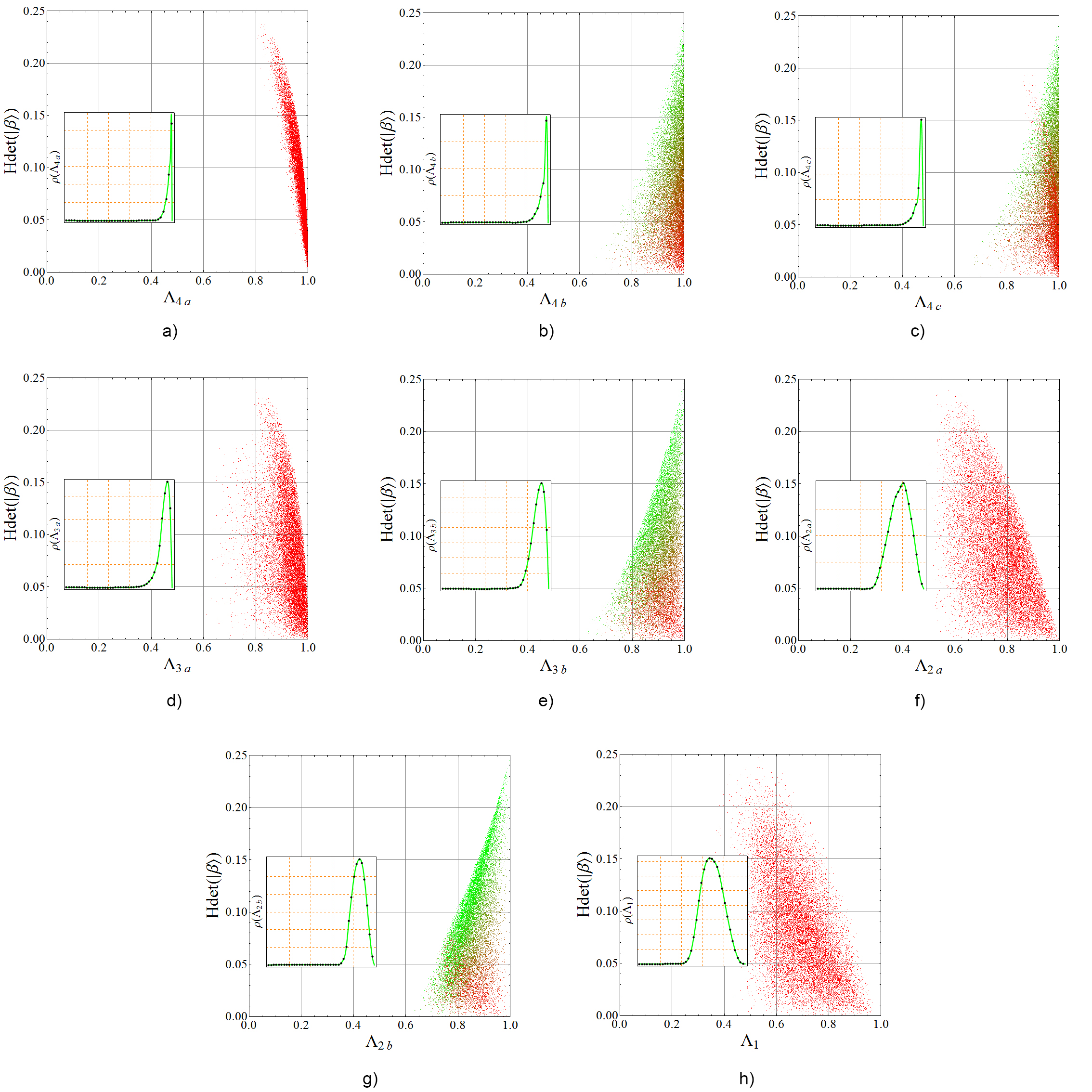}\hspace{1pc} 
\caption{\footnotesize Probability distribution of the maximum overlap $\Lambda_i$ for each of the Ac\'in classes 
%
%
(see Table 2) in the inset of each panel obtained (vertical scale on the left)
for an ensemble of $10^{5}$ three-qubit random states. Graphs of classes 4c and 4d are equivalent so this last was omitted (see details in the core text). The main plot represents the corresponding dispersion graph showing ${\rm Hdet}(\vert \beta \rangle)$ for each one (vertical scale on the right) and the value of ${\rm Hdet}(\vert \varphi \rangle)$, colored from red (separable) to green (maximal genuine entanglement).}
\label{fig8}
\end{figure}

Numerical results are showed jointly in Figure \ref{fig8}. First, the line plot shows the value of $\rho(\Lambda_i)$ on the left axis versus the value of projection $\Lambda_i$ in the horizontal axis. Superposed, a dispersion plot of the entire set of states being analyzed is shown in color. Each dot represents a random state located vertically on their projection value $\Lambda_i$ and horizontally in its  Hyperdeterminant value ${\rm Hdet}(\vert \beta \rangle)$, which remains invariant under the local optimization procedure. Additionally, each dot is colored in agreement with the Hyperdeterminant of the best class element $\vert \varphi \rangle$ obtained in the optimization. Colors are assigned from red for separable states to green for maximal genuine entanglement. This structure of the plot allows one to compare the closeness between $\vert \beta \rangle$ and $\vert \varphi \rangle$ in terms of genuine entanglement. Note the graph corresponding to class $4d$ has been omitted because its es equivalent to that of class $4c$: all coefficients in the class are real, then by exchanging $0$ and $1$ in all qubits and swapping the qubits $1$ and $3$ we get the same state with local operations. Thus, the maximal overlap and the hyperdeterminant statistics do not change.

Note particularly how in the Figure \ref{fig8}a the closest class states have ${\rm Hdet}(\vert \varphi \rangle)=0$ for some random states which have ${\rm Hdet}(\vert \beta \rangle)$ near from the highest value $\frac{1}{4}$ maintaining a closer distance $\Lambda_{4a} \approx 1$. The opposite phenomenon is also observed in Figures ~\ref{fig8}b, ~\ref{fig8}c, ~\ref{fig8}e and ~\ref{fig8}g where some class states with ${\rm Hdet}(\vert \varphi \rangle) \approx \frac{1}{4}$ (in green) are close to some random states with lower ${\rm Hdet}(\vert \beta \rangle)$ values. On the other hand, in Ref. \cite{Zyc05} the distribution of the fidelity between two random states has been computed analytically.  However in our case the problem becomes more complicated due to the optimization of the fidelity over all local unitaries.

\section{The entanglement polytope of three qubits}\label{entpoly}
Let $\lambda_k^{\min}$ denote the smallest eigenvalue of the reduced density matrix of the subsystem
of three qubits, where  $k=A,B, C$. The following set of compatibility conditions  
\begin{equation}\label{polyin}
\lambda_A^{\min}\le \lambda_B^{\min}+\lambda_C^{\min}, \quad \lambda_B^{\min}\le \lambda_A^{\min}+\lambda_C^{\min}, \quad \lambda_C^{\min}\le \lambda_A^{\min}+\lambda_B^{\min}.
\end{equation}
form  particular examples of {\it polygon inequalities} obtained by Higuchi {\it et. al.} for systems of several qubits \cite{Hig03}. The smaller eigenvalue of a one-qubit system is not larger then $1/2$ 
so that  $0\le\lambda_k^{\min}\le 1/2$. Inequalities (\ref{polyin}) determine jointly a convex polytope 
in the three-space $( \lambda_A^{\min}, \lambda_B^{\min}, \lambda_C^{\min})$.
Its five vertices represent distinguished three-qubit states: 
 fully separable states are identified by the point $(0,0,0)$ 
whereas points $(1/2,1/2,0)$, $(1/2,0,1/2)$ and $(0,1/2,1/2)$ stand for bi-separable states. 
The $GHZ$-state is located at $(1/2,1/2,1/2)$. The convex hull of these points is known as the {\it Kirwan polytope} \cite{SOK12,Wal13,SOK14}. 
In addition, the identification of a state belonging to an entanglement classes 
reported in \cite{Han04} is summarized in Table \ref{tab1}.

\begin{figure}[h] 
\begin{tabular}{cccc}
\includegraphics[scale=0.55]{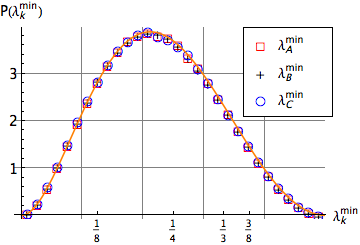} & & &
\includegraphics[scale=0.35]{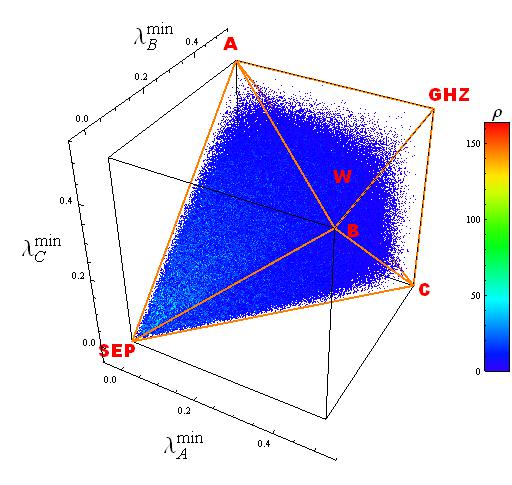}\\
(a) & & & (b)\\
\includegraphics[scale=0.35]{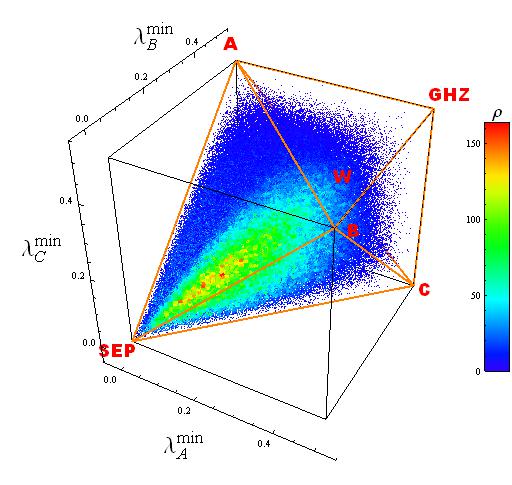} & & &
\includegraphics[scale=0.33]{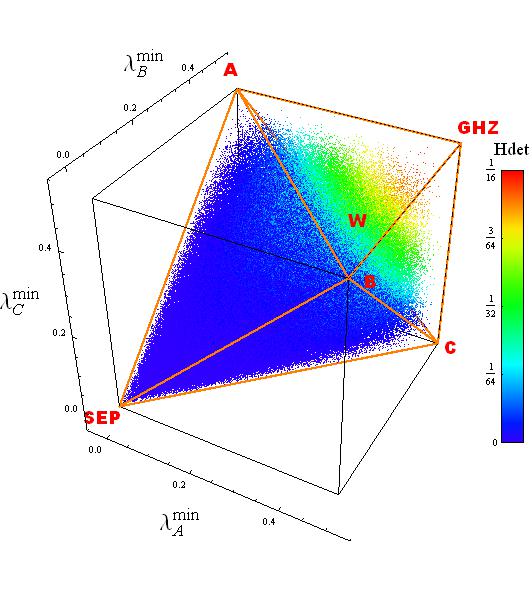}\\
(c) & & & (d)
\end{tabular}
\caption{\footnotesize (a) Probability distribution of the minimal eigenvalue 
of a single qubit reduced system (\ref{lmin}). (b) An ensemble $10^6$ of three-qubit random states depicted in the entanglement polytope. The color scale stands for the joint probability distribution. (c) Detail of (b): a transversal section by the plane which contains the points $SEP$, $C$ and $GHZ$. (d) The ensemble of three qubit random states labeled by colour set according to the value of  the invariant $I_6$.}

\label{poly3q}
\end{figure}

Consider now an ensemble of three-qubit random states. For such states the probability distribution of the minimal eigenvalue of a single-particle reduced density matrix fulfills
$P(\lambda_{\min})=P(\lambda_A^{\min})=P(\lambda_B^{\min})=P(\lambda_C^{\min})$. Using the following relation between the two eigenvalues $\vartheta_1$ and $\vartheta_2$ of a single qubit reduced density matrix
\begin{equation*}
\lambda_{\min}=\min(\vartheta_1,\vartheta_2)=\frac12(\vartheta_1+\vartheta_2)-\frac12 \vert \vartheta_1-\vartheta_2\vert,
\end{equation*}
we can compute the probability distribution of the minimal eigenvalue $\lambda_{\min}$ as
\begin{equation}
 P(\lambda_{\min})=\int_0^1\int_0^1 d\vartheta_1 d\vartheta_2 {\cal P}(\vartheta_1,\vartheta_2) \delta [\lambda_{\min}-(\vartheta_1+\vartheta_2)/2 +\vert \vartheta_1-\vartheta_2\vert/2)],
\end{equation}
where ${\cal P}(\vartheta_1,\vartheta_2)$ is the joint density (\ref{plambdas}) and $\delta$ stands for the Dirac delta function. Performing the integral we obtain
\begin{equation}\label{lmin}
 P(\lambda_{\min})=420 [\lambda_{\min}(2 \lambda_{\min}-1)(1-\lambda_{\min})]^2, \quad 0\le \lambda_{\min} \le 1/2.
\end{equation}
This distribution is depicted in Fig.\,\ref{poly3q}. Besides, a direct calculation yields the average value $\langle \lambda_{\min}\rangle=29/128$. In general, the $k$-the moment of $\lambda_{\min}$ reads
\begin{equation}
 \langle \lambda_{\min}^k\rangle= \frac{105}{2^k}\left[\frac{\Gamma(k+3)}{\Gamma(k+6)}-\frac{\Gamma(k+4)}{\Gamma(k+7)}+\frac{\Gamma(k+5)}{4\Gamma(k+8)}\right].
\end{equation}
Note that a  given pure state can be identified with a point in the entanglement polytope. Its coordinates are $(\lambda_A^{\min},\lambda_B^{\min},\lambda_C^{\min})$. This is shown in Fig.~\ref{poly3q}(b)
for an ensemble of $10^6$ three-qubit random states colored according to their joint probability distribution in the polytope. 
To compute such probability distribution the space containing the whole polytope $[0,\frac{1}{2}]^{\times 3}$ was divided in $80^3$ cubic cells. Then, we state the statistics of random states falling in each cell to get the probability density of those states (by volume unity). 
Note that the closer the points are to the faces the lower the value of the distribution. In Fig.~\ref{poly3q}(c) we depict a transverse cut by the plane containing the vertices $SEP$, $C$ and $GHZ$ to depict the distribution of the inner points. This shows that random states are more concentrated near the line joining the vertices $SEP$ and $GHZ$, which corresponds to class 2a. 

On the other hand, two quantum pure states attain the same amount of entanglement if they belong to the same class, that is to say if there is a finite probability of success that they can be converted into each other using stochastic local operations and classical communication, referred as SLOCC by its acronyms. For the case of three qubits, there exist two SLOCC classes of entanglement: the one containing the $GHZ$ state, which exhibits genuine entanglement and the $W$ class \cite{dur}.
These classes can be distinguished from the entanglement polytope. Numerical calculation shows that around 6\% of the states are placed in the upper polytope, so that they belong to the GHZ SLOCC class \cite{Wal13}. As the invariant $I_6$ discriminates between such classes in panel ~\ref{poly3q}(d) we show the ensemble of random states colored with respect to this invariant. For states placed near the bi-separable faces $I_6$
goes to zero, whereas the states landing in the GHZ simplex are characterized by a positive  value of 
this invariant. An equivalent approach can be done dealing the maximum eigenvalues of the reduced single qubit density matrices. For such a case the joint probability distribution is known \cite{Chr14} and hence the fraction of random states in the GHZ pyramid was computed in Ref. \cite{Zha17} yielding $13/216\approx 6.02\%$ which is consistent with our numerical calculation.

%

\section{Conclusions and future work}\label{conclu}

We analyzed the probability distributions of the six parameters 
determining the five-terms decomposition (\ref{acin1}) of a random three-qubit state.
The phase of the complex coefficient occurs to be uniformly distributed. The distributions of the amplitudes $\lambda_0$ and $\lambda_4$ differ from the distribution describing the remaining three coefficients.
Interestingly, these two coefficients can be related with the degree of entanglement as the invariant $J_4$ depends only on them. In addition, we have also analyzed the probability distributions of two sets of polynomial invariants. The invariants $I_1, I_2$ and $I_3$ follow the same distribution.
Thus, out of the five independent invariants only three are necessary to characterize entanglement in three-qubit states. This fact is consistent with the second set of invariants reported by Ac\'in {\it et al.} as the distributions of the invariants $J_1, J_2$ and $J_3$ do coincide. For each invariant its 
mean value 
was computed using symbolic integration with respect to  the unitary invariant Haar measure.

On the other hand, the set of invariants $\{J_k\}$ allows us to identify certain entanglement classes, 
whose entanglement was described through the minimal decomposition entropy. Moreover, highly entangled states with respect this measure were identified in each class. Our results imply 
that the more terms in the decomposition (\ref{acin1})
of a three-qubit state, the larger its degree of entanglement 
measured by the minimal decomposition entropy. 

In this sense the use of RIU entropy as an exemplary measure of quantum entanglement 
allows us to provide a classification of three qubit states and to describe their hierarchy. 
The invariants with respect to local transformations are useful to identify certain types of entangled structures in the entire system. As shown in Fig. \ref{fig8}
the states displaying genuine entanglement appear closer from other 
states in the classes with not genuine entanglement. 
Although smooth measures of entanglement depend 
on the state in a continuous way, 
a small variation of a state can lead to a considerable change of its entanglement.
This feature was observed in larger systems \cite{delgadoAT}.

In such a scenario, the current analysis in the quest of understanding the hierarchy of entanglement, 
could set directions to transform states from maximally entangled into separable ones. 
%
%
Recently, using the $SU(2)$ decomposition procedure \cite{delgadoSU2a, delgadoSU2b}, has been clear the existence of basic $U(1) \times SU(2)$ operations among entangled pairs, showing how the entanglement phenomena can be generated in a structured way transiting from  separable to genuine entangled states. It suggests that programmed local operations combined with 2-entangling operations (those entangling two previous entangled pairs) can be realized in order to connect such state types. Thus, basic separable states could be transformed into maximal entangled states as $\vert GHZ\rangle$ and $\vert W\rangle$ only with a series of such operations. In a more ambitious task, those single types of operations suggest they could be responsible for the transit from certain classes in other among the hierarchies of entanglement.  In such process, the track in the change of the invariants values could to provide a strong road-map for such transit.

Finally, we have analyzed the probability distribution of the maximal fidelity of a random state with 
respect to the closest representative of each entanglement class. 
The highest maximal fidelity is obtained for classes 4a-4d listed in Table 2. 
%
This can be seen from the fact that the five Ac\'in coefficients in each random state are in average non-trivial.
%
%
We hope our results shed some light on the non-local properties of three-qubit pure random states.

\section*{Acknowledgments}

The support of Escuela de Ingenier\'ia y Ciencias of Tecnol\'ogico de Monterrey as well as the support of
CONACyT are gratefully acknowledged.
K{\.Z} acknowledges support by Narodowe Centrum Nauki
under the grant number DEC-2015/18/A/ST2/00274.


\end{document}